\documentclass[12pt]{iopart}
\usepackage{iopams}
\usepackage{graphicx}
\usepackage{amssymb}
\usepackage{cite}
\usepackage{mathrsfs}
\begin{document}

\title{Star product and ordered moments of photon creation and annihilation operators}

\author{S N Filippov$^1$ and V I Man'ko$^{1,2}$}

\address{$^1$ Moscow Institute of Physics and Technology, Dolgoprudnyi, Moscow Region, Russia\\
$^2$ P.~N.~Lebedev Physical Institute, Russian Academy of
Sciences, Moscow, Russia}

\eads{\mailto{sergey.filippov@phystech.edu},
\mailto{manko@sci.lebedev.ru}}

\begin{abstract}
We develop a star-product scheme of symbols defined by the
normally ordered powers of the creation and annihilation photon
operators, $(\hat{a}^{\dag})^m \hat{a}^n$. The corresponding phase
space is a two-dimensional lattice with nodes $(m,n)$ given by
pairs of nonnegative integers. The star-product kernel of symbols
on the lattice and intertwining kernels to other schemes are found
in explicit form. Analysis of peculiar properties of the
star-product kernel results in new sum relations for factorials.
Advantages of the developed star-product scheme for describing
dynamics of quantum systems are discussed and time evolution
equations in terms of the ordered moments are derived.
\end{abstract}

\maketitle

\section{\label{introduction} Introduction}
This article concerns quantum states of one-mode radiation field
and their time development by using normally ordered moments
$\langle (\hat{a}^{\dag})^m \hat{a}^n \rangle$ of photon creation
and annihilation operators, $\hat{a}^{\dag}$ and $\hat{a}$.
Ordering of creation and annihilation operators was studied
extensively in the 1960s in connection with the quasiprobability
distributions (see, e.g.,
\cite{klauder-sudarshan,cahill-glauber-1,cahill-glauber-2}). Thus,
the Husimi-Kano function $Q(\alpha)$~\cite{husimi,kano} is closely
related with the normally ordered density operator
$\hat{\rho}=\sum_{m,n}\rho_{mn}^{(\rm n)} (\hat{a}^{\dag})^m
\hat{a}^n$, namely, $Q(\alpha) = \sum_{m,n}\rho_{mn}^{(\rm n)}
(\alpha^{\ast})^m {\alpha}^n$, where $\alpha^{\ast}$ denotes the
complex conjugate of $\alpha$. Similarly, the Wigner
$W$-quasidistribution~\cite{wigner} and the Sudarshan-Glauber
$P$-quasidistribution~\cite{sudarshan,glauber} are expressed
through symmetrically and antinormally ordered representations of
the density operator, respectively (see,
e.g.,~\cite{lee,schleich}). The quasidistributions on phase space
do not limit to $Q$, $W$, and $P$-functions and are reviewed in
several papers (see, e.g.,~\cite{lee,hillery,zachos}). The recent
detailed review of the phase-space approach is presented by
Vourdas~\cite{vourdas}. An advantage of the phase-space
description of quantum states is that one deals with functions
($\mathbb{C} \mapsto \mathbb{R}$) instead of operators. As
equations for quasidistributions do not involve any operator, they
are sometimes easier to handle than the Schr\"{o}dinger or von
Neumann equations~\cite{lee}. Moreover, in the phase-space
formalism, quantum phenomena are known to be interpreted in the
classical-like manner~\cite{moyal}.

Being a rather good alternative to the density operator, the
quasidistribution functions $Q$, $W$, and $P$ are all equivalent
in the sense that they all contain the thorough physically
meaningful information about the state~\cite{lee}. As the
quasidistributions are equivalent, the question arises itself: why
do not we use only one of them for all the problems? The answer
lies in actual applications of the quasidistributions. For
example, using the classical interpretation of the Wigner
$W$-function in collision problems, it is possible to make some
reasonable approximations and obtain an appropriate but still
accurate solution to the problem without requiring excessive
computer time and expense. On the other hand, the nonnegativity
and smoothness of the Husimi-Kano $Q$-function make it
advantageous for the analysis of classically chaotic nonlinear
systems~\cite{lee}. Also, the $Q$-function of the radiation field
can be measured at optical frequencies via an eight-port homodyne
detection scheme (see, e.g.,~\cite{schleich}). Finally, the
negativity of the Sudarshan-Glauber $P$-function is a commonly
accepted criterion of the state's non-classicality, however, to
observe the fact $P(\alpha)<0$ is not that easy (see,
e.g.,~\cite{vogel-00,kiesel-10}). We can draw a conclusion that
although all the quasidistribution functions are equivalent, the
different functions exhibit different properties and obey
different dynamical equations. Those differences specify the most
advantageous representation for a particular problem.

The set of normally ordered moments $\langle (\hat{a}^{\dag})^m
\hat{a}^n \rangle$ can also be treated as a quasiprobability
function of two nonnegative integers, i.e. $f: \mathbb{Z}_{+}
\times \mathbb{Z}_{+} \mapsto \mathbb{C}$. An interest to the
ordered moments rose in the 1990s and was encouraged by the
advances of measuring quantum states of light. Similar to the
conventional quasidistributions, the moments $\langle
(\hat{a}^{\dag})^m \hat{a}^n \rangle$ thoroughly determine the
density operator $\hat{\rho}$ and the explicit relation is found
in the
papers~\cite{wunsche-90,lee-ct,herzog,wunsche-buzek,wunsche-99}.
Despite the moments $\langle (\hat{a}^{\dag})^m \hat{a}^n \rangle$
are but one of plenty equivalent prescriptions to deal with
quantum states (see, e.g.,~\cite{vourdas}), there is a particular
problem where the moments play an indispensable role: a
measurement of itinerant microwave quantum states.

The matter is that at optical wavelengths there is no need to
calculate the normally ordered moments and then use them to find
the density operator. The output of the homodyne detection scheme
is rotated quadrature distributions $w(X,\theta)$,
$X\in\mathbb{R}$, $\theta\in[0,\pi]$, also referred to as optical
tomogram (see, e.g.,~\cite{raymer,schiller,porzio} and the review
\cite{lvovsky}). There exists an explicit formula for
reconstructing the density operator $\hat{\rho}$ in terms of the
optical tomogram (see,
e.g.,~\cite{bertrand,vogel-risken,mancini-manko-tombesi}). The
normally ordered moments are easily expressed through the measured
optical tomogram as well~\cite{richter,wunsche-96}.

At microwave frequencies, conversely, to measure the rotated
quadrature distributions $w(X,\theta)$ is a challenge
\cite{mallet} because of the strong thermal noise added and the
absence of single-photon detectors in such spectral region.
However, it has been reported recently how to extract the
lower-order moments $\langle (\hat{a}^{\dag})^m \hat{a}^n \rangle$
of microwave quantum states themselves and exclude the noise
contribution to experimental
data~\cite{menzel,mariantoni,eichler}. For details we refer the
reader to the paper~\cite{filippov}. Thus, in the microwave
domain, the lower-order moments are experimentally determined and
contain the primary information about a quantum state.

Given only normally ordered moments of the microwave radiation
field, it is reasonable to associate a quantum state with the
measurable quasiprobability function $f(m,n)\equiv\langle
(\hat{a}^{\dag})^m \hat{a}^n \rangle$ of two nonnegative integers
and consider the time evolution $f(m,n;t)$ governed by a
Hamiltonian $\hat{H}$ in presence of decoherence processes. This
motivates us to follow a star-product
approach~\cite{stratonovich,berezin,garcia-bondia,brif-mann-jpa,brif-mann-pra,oman'ko-JPA,oman'ko-vitale}
and develop the particular star-product scheme on a
two-dimensional lattice $(m,n)$, where any operator $\hat{A}$ is
associated with a symbol $f_A(m,n)={\rm Tr} \big[
(\hat{a}^{\dag})^m \hat{a}^n \hat{A} \big]$ and the product
$\hat{A}\hat{B}$ of two operators corresponds to a star product
(also known as twisted product) of the corresponding symbols, i.e.
$f_{AB}(m,n) \equiv [f_A \star f_B](m,n)$. Hence, one can
associate the density operator $\hat{\rho}$ of a quantum state
with the normally ordered moments $\langle (\hat{a}^{\dag})^m
\hat{a}^n \rangle \equiv f_{\rho}(m,n)$ and the Hamiltonian
$\hat{H}$ with its symbol ${\rm Tr} \big[ (\hat{a}^{\dag})^m
\hat{a}^n \hat{H} \big] \equiv f_{H}(m,n)$. Then the unitary
evolution of a quantum state can be easily written in terms of the
introduced symbols as
\begin{equation}
\label{unitary-evolution}\frac{\partial f_{\rho}(m,n)}{\partial t}
= -\frac{i}{\hbar} \left[ f_H \star f_{\rho} - f_{\rho} \star f_H
\right](m,n), \qquad m,n=0,1,\ldots,
\end{equation}

\noindent where $\hbar$ is the Planck constant.

Formula (\ref{unitary-evolution}) is nothing else but the
evolution equation for the normally ordered moments. Therefore,
the time development of a quantum state is expressed through the
measurable characteristics, which resembles the expectation value
approach~\cite{weigert-1,weigert-2} and the
tomographic-probability
approach~\cite{mancini-manko-tombesi,mancini-manko-tombesi-found-phys,ibort}
to quantum mechanics. According to the latter one, the measurable
tomographic probability is a primary object to describe quantum
states. The evolution equations for tomograms were found, e.g., in
\cite{omanko-jrlr,moshinsky,korennoy}.

A kernel that determines the star product $\star$ in formula
(\ref{unitary-evolution}) is not known in the literature and is
expressed through factorials in \Sref{sec-Star-Product-Kernel}.
Since the kernel of any star product scheme is to satisfy a
nontrivial sum relation, we derive a new relation on factorials in
passing.

Let us note that a unitary evolution of any quasidistribution
function can be written in the form of Eq.
(\ref{unitary-evolution}) (see, e.g.,~\cite{lee}) but the
specificity of our particular case is that the symbol $f_{H}(m,n)$
is not determined even for the harmonic oscillator Hamiltonian
$\hat{H} = \hbar \omega (\hat{a}^{\dag} \hat{a} + 1/2)$ because
the trace is diverging. Nevertheless, in \Sref{section-evolution}
we show that the symbols $[ f_H \star f_{\rho}](m,n)$ and
$[f_{\rho} \star f_H ](m,n)$ can still be determined and
Eq.~(\ref{unitary-evolution}) takes the form of a difference
equation for $f_{\rho}(m,n)$.

The aim of this paper is to develop the star-product scheme of
symbols given by the normally ordered moments, to explore
properties of the star-product kernel, and to derive unknown
evolution equations for the moments via the star-product approach.
It is worth pointing out that we study the star-product scheme not
of functions on the conventional phase space $(q,p)$ but of
functions on the two-dimensional lattice $(m,n)$, where $m,n$ are
nonnegative integers that determine the moments $\langle
(\hat{a}^{\dag})^m \hat{a}^n \rangle$. This peculiarity provides a
nontrivial change of variables and, as we will see, yields the
difference quantum basic equations in contrast to the partial
differential equations for the quasidistributions $Q(\alpha)$,
$W(\alpha)$, and $P(\alpha)$.

The article is organized as follows.

In \Sref{section-symbols}, we show how to treat operators as
symbols defined by the normally ordered powers of creation and
annihilation operators. Also, we revisit the reconstruction of
operators given by their symbols, i.e. the problem of moments. In
\Sref{sec-Star-Product-Kernel}, we develop the star-product
formalism~\cite{oman'ko-JPA,oman'ko-vitale} of the introduced
symbols. In particular, we find the star-product kernel and
intertwining kernels. We calculate those kernels in the explicit
form and use their particular properties to derive new sum
relations involving factorials. In \Sref{section-evolution}, we
apply the developed star-product scheme to derive the time
evolution of moments. Unitary and non-unitary evolutions are
considered. Conclusions are presented in
\Sref{section-conclusions}.

\section{\label{section-symbols} Quantization scheme: symbols of operators}
Unless otherwise stated, we assume that the expression ${\rm
Tr}[\cdot]$ is well defined. Thus, we intensionally avoid
discussions of the convergence problems and focus our attention on
the recipe to deal with symbols of operators.

\begin{figure}
\begin{center}
\includegraphics{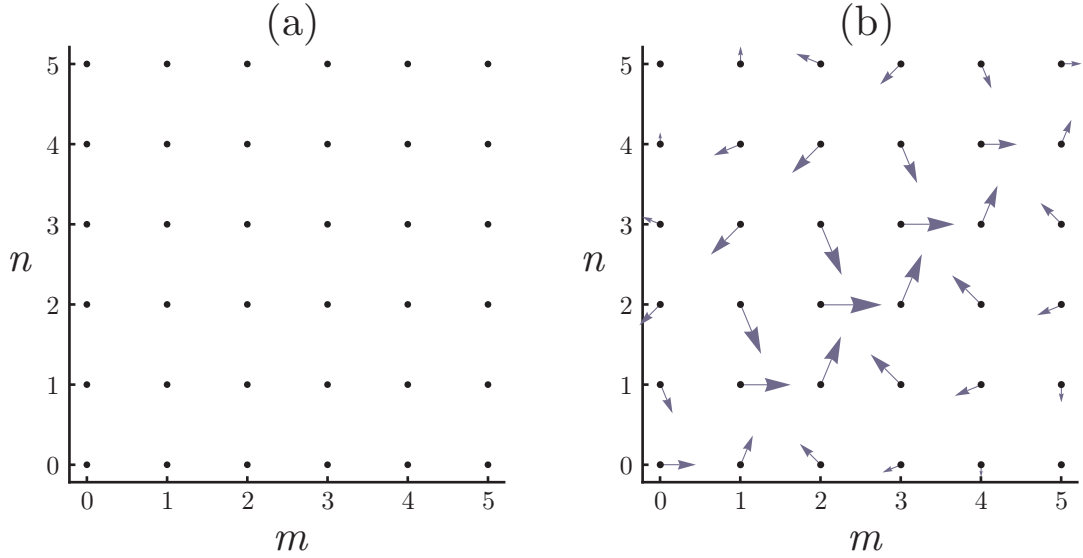}
\caption{\label{figure-phase-space} (a) Phase space is a lattice
whose nodes are labelled by two nonnegative integers $(m,n)$.
Symbol $f_{\rho}(m,n)$ is a complex valued function on the
lattice. (b) Vectors $({\rm Re} f_{\rho}(m,n)/m!n!; {\rm Im}
f_{\rho}(m,n)/m!n!)$ represent complex values of the symbol
$f_{\rho}(m,n)$ of the squeezed vacuum pure state $\rho$ with
squeezing parameter $r=1.3e^{i 5\pi/4}$.}
\end{center}
\end{figure}

Consider an operator $\hat{A}$ acting on the same Hilbert space as
the density operator $\hat{\rho}$. Assuming the existence of the
trace
\begin{equation}
\label{dequantization} f_{A}(m,n) \equiv {\rm Tr} \big[
(\hat{a}^{\dag})^m \hat{a}^n \hat{A} \big] \equiv {\rm Tr} \big[
\hat{U}^{\dag}(m,n) \hat{A} \big],
\end{equation}

\noindent we will refer to the function $f_{A}(m,n)$ of two
discrete variables $m,n=0,1,\ldots$ as symbol of the operator
$\hat{A}$. For the sake of convenience, we have introduced the
dequantizer operator $\hat{U}(m,n) \equiv (\hat{a}^{\dag})^n
\hat{a}^m$. Symbols $f_{A}(m,n)$ are defined on the
two-dimensional lattice $(m,n)$ (Fig.~\ref{figure-phase-space}a).
An example of the symbol $f_{\rho}(m,n)$ is illustrated for a
squeezed vacuum state $\rho$ in Fig.~\ref{figure-phase-space}b.

The set $\{f_{A}(m,n)\}_{m,n=0}^{\infty}$ is known to be
informationally complete
\cite{wunsche-90,lee-ct,herzog,wunsche-buzek,wunsche-99}. In other
words, the operator $\hat{A}$ can be reconstructed as follows:
\begin{eqnarray}
\label{quantization-1} && \hat{A} = \sum_{m,n=0}^{\infty}
f_{A}(m,n) \hat{D}(m,n),
\\
\label{quantization-2} && \hat{D}(m,n) = \frac{1}{m!n!}
\sum_{j=-\{m,n\}}^{\infty} \frac{(-1)^j (m+n+j)!}{(m+j)!(n+j)!}
(\hat{a}^{\dag})^{n+j} \hat{a}^{m+j},
\end{eqnarray}

\noindent where $\{m,n\} = \min(m,n)$ and the operator
$\hat{D}(m,n)$ is called quantizer.

Using the Fock state representation of powers of creation and
annihilation operators
\begin{equation}
\fl (\hat{a}^{\dag})^m = \sum_{k=0}^{\infty}
\sqrt{\frac{(m+k)!}{k!}} |m+k\rangle \langle k|, \qquad \hat{a}^n
= \sum_{k=0}^{\infty} \sqrt{\frac{(n+k)!}{k!}} |k\rangle \langle
n+k|,
\end{equation}

\noindent the dequantizer and quantizer can be rewritten in the
form
\begin{eqnarray}
\label{U-normal-Fock} && \hat{U}(m,n) = \sum_{k=0}^{\infty}
\frac{\sqrt{(n+k)!(m+k)!}}{k!} |n+k\rangle
\langle m+k|, \\
\label{D-normal-Fock} && \hat{D}(m,n) = \sum_{j=0}^{\{m,n\}}
\frac{(-1)^j |n-j\rangle \langle m-j|}{j! \sqrt{(m-j)!(n-j)!}}.
\end{eqnarray}

The dequantizer and quantizer are shown to be orthogonal in the
sense of trace operation
\cite{wunsche-90,lee-ct,herzog,wunsche-buzek,wunsche-99}
\begin{equation}
\label{delta} {\rm Tr} \big[ \hat{U}^{\dag}(m,n) \hat{D}(m',n')
\big] = \delta_{m,m'} \delta_{n,n'},
\end{equation}

\noindent where $\delta_{i,j}$ is the conventional Kronecker
delta-symbol.

Let us now check that Eqs.
(\ref{U-normal-Fock})--(\ref{D-normal-Fock}) define an
informationally complete scheme. It is shown in the
papers~\cite{manko-marmo-simoni-etal,manko-marmo-simoni-vent,mms-sudarshan-vent}
that a quantization scheme is informationally complete
(tomographic) if and only if
\begin{equation}
\label{tomogr-s-p-require} \sum_{m,n=0}^{\infty} || D(m,n)
\rangle\rangle \langle\langle U(m,n) || = \hat{\mathcal{I}},
\end{equation}

\noindent where $\hat{\mathcal{I}}$ is the identity
super-operator, $|| D(m,n) \rangle\rangle$ and $|| U(m,n)
\rangle\rangle$ are vectors constructed from the quantizer and the
dequantizer by the procedure of representing matrices as vectors.
In our case we have
\begin{eqnarray}
&& || U(m,n) \rangle\rangle = \sum_{k=0}^{\infty}
\frac{\sqrt{(n+k)!(m+k)!}}{k!} |n+k\rangle \otimes |m+k \rangle,
\\
&& || D(m,n) \rangle\rangle = \sum_{j=0}^{\{m,n\}} \frac{(-1)^j
|n-j\rangle \otimes |m-j \rangle}{j! \sqrt{(m-j)!(n-j)!}}.
\end{eqnarray}

Direct calculation yields
\begin{eqnarray}
&& \langle p | \otimes \langle q | \left( \sum_{m,n=0}^{\infty} ||
D(m,n) \rangle\rangle \langle\langle U(m,n) || \right) | r \rangle
\otimes | s \rangle
\nonumber\\
&& = \sqrt{\frac{r!s!}{q!p!}} \delta_{p+s,q+r} \sum_{n=p}^{r}
\frac{(-1)^{n-p}}{(r-n)!(n-p)!} \nonumber\\
&& = \sqrt{\frac{r!s!}{q!p!}} \delta_{p+s,q+r}
\frac{(1-1)^{r-p}}{(r-p)!}= \sqrt{\frac{r!s!}{q!p!}}
\delta_{p+s,q+r} \delta_{p,r} = \delta_{p,r} \delta_{q,s},
\end{eqnarray}

\noindent that is the relation (\ref{tomogr-s-p-require}) holds
true indeed.

\section{\label{sec-Star-Product-Kernel} Star product}
By definition, a star product of symbols $f_A$ and $f_B$ is
nothing else but the symbol $f_{AB}(m,n)$ of the product of two
operators $\hat{A}$ and $\hat{B}$, namely,
\begin{eqnarray}
\label{star-product} && (f_A \star f_B) (m,n) \equiv f_{AB}(m,n)
\nonumber\\
&& = \sum_{m',n',m'',n''=0}^{\infty} f_A(m',n') f_B(m'',n'')
K(m,n;m',n';m'',n''),
\end{eqnarray}

\noindent where the kernel $K$ is expressed in terms of the
dequantizer and quantizer operators as follows:
\begin{equation}
\label{kernel} \fl K(m,n;m',n';m'',n'') = {\rm Tr} \big[
\hat{U}^{\dag}(m,n) \hat{D}(m',n') \hat{D}(m'',n'') \big].
\end{equation}

\noindent From definition (\ref{star-product}) it follows
immediately that the star product is associative and the
star-product kernel satisfies a nontrivial relation
\begin{eqnarray}
\label{KK-KK} && \sum_{k,l=0}^{\infty} K(m,n;k,l;m''',n''')
K(k,l;m',n';m'',n'') \nonumber\\
&& = \sum_{k,l=0}^{\infty} K(m,n;m',n';k,l)
K(k,l;m'',n'';m''',n'''),
\end{eqnarray}

\noindent which is a consequence of the relation $f_A \star f_B
\star f_C = (f_A \star f_B) \star f_C = f_A \star (f_B \star
f_C)$.

Substituting (\ref{U-normal-Fock}) and (\ref{D-normal-Fock}) for
the dequantizer and quantizer in (\ref{kernel}), after some
algebra, we obtain the star-product kernel in the explicit form
\begin{eqnarray}
\label{kernel-explicit} && K(m,n;m',n';m'',n'')
\nonumber\\
&& = \frac{(-1)^{m'-n''} (m'+m''-m)!}{m'!n''!(m''-m)!(n'-n)!}
\delta_{m+n'+n'',n+m'+m''} \nonumber\\
&& = \frac{(-1)^{m'-n''} \Gamma(m'+m''-m + 1)
\delta_{m+n'+n'',n+m'+m''} }{\Gamma(m'+1) \Gamma(n''+1)
\Gamma(m''-m+1) \Gamma(n'-n+1)},
\end{eqnarray}

\noindent where $\Gamma$ is the conventional Euler gamma function.

\begin{figure}
\begin{center}
\includegraphics{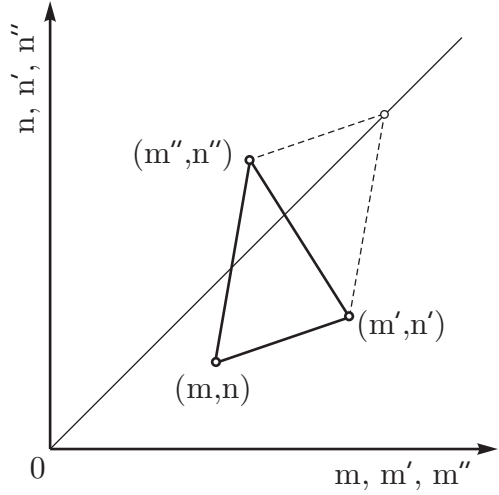}
\caption{\label{figure} The kernel $K(m,n;m',n';m'',n'')$ is
nonzero if the points $(m,n)$, $(m',n')$, and $(m'',n'')$
determine a parallelogram such that its forth vertex, that is
opposite to $(m,n)$, belongs to a bisecting line. Also the
conditions $m \le m''$, $n \le n'$ are to be met.}
\end{center}
\end{figure}

Using the property of the Kronecker delta-symbol, the kernel can
also be rewritten as follows:
\begin{equation}
\label{kernel-explicit-F} \fl K(m,n;m',n';m'',n'') = F(m',m''-m)
F(n'',n'-n) \delta_{m+n'+n'',n+m'+m''},
\end{equation}

\noindent where $F(a,b) = (-1)^{a} \sqrt{(a+b)!} / a!b!$. In
Fig.~\ref{figure} we illustrate the restrictions on arguments of
the kernel (\ref{kernel-explicit}) under which it can take
non-zero values.

Let us now consider how the nontrivial relation (\ref{KK-KK})
looks like for the kernel (\ref{kernel-explicit}). Substituting
(\ref{kernel-explicit-F}) for $K$ in (\ref{KK-KK}), we obtain the
equality
\begin{eqnarray}
&& \fl \sum_{k} F(k,m'''-m) F(m',m''-k) F(n''',k+m'''-m-n''')
F(n'',m'+m''-n''-k) \nonumber\\
&& \fl = \sum_{l} F(m',l+n'-n-m') F(l,n'-n) F(m'',n''+n'''-m''-l)
F(n''',n''-l)
\end{eqnarray}

\noindent which is to hold true whenever $m'+m''+m'''+n =
n'+n''+n'''+m$. Expressing $n'$ through the other variables and
using the explicit form of the function $F$ yields
\begin{eqnarray}
\label{sum-rel-1} \fl && \sum_{l=\lceil 0,n''+n'''+m-m''-m'''
\rceil}^{\{n''+n'''-m'',n''\}} \frac{(-1)^l
(l+m'+m''+m'''-n''-n'''-m)!(n''+n'''-l)!}{l!(l+m''+m'''-n''-n'''-m)!(n''+n'''-m''-l)!
(n''-l)!} \nonumber\\
\fl && = \frac{(-1)^{n''+m''}
m''!(m'+m''+m'''-n''-n'''-m)!}{(m'''-m)!n''!}\nonumber\\
\fl && \quad \times \sum_{k=\lceil 0, m+n'''+m'''
\rceil}^{\{m'',m'+m''-n''\}} \frac{(-1)^k
(k+m'''-m)!(m'+m''-k)!}{k!(k-m-n'''-m''')!(m''-k)!(m'+m''-n''-k)!},
\end{eqnarray}

\noindent where $\lceil a,b \rceil = \max(a,b)$ and $\{ a,b \} =
\min(a,b)$. Although the obtained relation is rather complicated,
it is valid for all non-negative integers $m$, $m'$, $m''$,
$m'''$, $n$, $n''$, $n'''$. For instance, if we put
$m'+m''-n''=0$, then the summation over $k$ in the right-hand side
of Eq. (\ref{sum-rel-1}) reduces to a single term $k=0$ provided
$m+n'''-m''' \le 0$. As a result we derive a new property
\begin{equation}
\label{KK-KK-1-simple} \sum_{l=\lceil 0,m'-M
\rceil}^{\{n'''+m',n''\}} \frac{(-1)^l (M+l)! (n''+n'''-l)! }{l!
(M-m'+l)! (n'''+m'-l)! (n''-l)!} = (-1)^{m'},
\end{equation}

\noindent where $M=m'''-m-n''' \ge 0$ and $n'' \ge m'$. Note that
the result of summation does not depend on $m'',n'',n'''$. In
particular, if we choose $n'''=0$, then
\begin{equation}
\sum_{l=\lceil 0,m'-M \rceil}^{m'} \frac{(-1)^l
(M+l)!}{l!(M-m'+l)!(m'-l)!} = (-1)^{m'}.
\end{equation}

\noindent Similarly, if we fix $m'=0$ in Eq.
(\ref{KK-KK-1-simple}), then
\begin{equation}
\label{sum-rel-4} \sum_{l=0}^{\{n'',n'''\}} \frac{(-1)^l
(n''+n'''-l)!}{l!(n''-l)!(n'''-l)!} = 1.
\end{equation}

\subsection{\label{subsec-Interteining-Kernels} Intertwining kernel
between star-product schemes}

Along with the considered star-product scheme with dequantizers
(\ref{U-normal-Fock}) and quantizers (\ref{D-normal-Fock}), there
exist many other star-product schemes. In order to distinguish
them let us use superscripts $(\mathfrak{N})$ for the just
developed scheme of the normally ordered moments and write
$\hat{U}^{(\mathfrak{N})}(m,n)$ and
$\hat{D}^{(\mathfrak{N})}(m,n)$.

As an example of the other quantization on the lattice $(m,n)$ we
may consider the star-product scheme with identical dequantizers
and quantizers $\hat{U}^{(\mathfrak{F})}(m',n') =
\hat{D}^{(\mathfrak{F})}(m',n') = | m' \rangle \langle n' |$. The
symbols, that correspond to the two different schemes, are related
by virtue of formulas
\begin{eqnarray}
\label{intertw-symbols} && f_A^{(\mathfrak{N})}(m,n) =
\sum_{m',n'=0}^{\infty} K_{\mathfrak{F} \rightarrow \mathfrak{N}}
(m,n;m',n') f_A^{(\mathfrak{F})}(m',n'), \\
&& f_A^{(\mathfrak{F})}(m',n') = \sum_{m,n=0}^{\infty} K_{
\mathfrak{N} \rightarrow \mathfrak{F}} (m',n';m,n)
f_A^{(\mathfrak{N})}(m,n),
\end{eqnarray}
\noindent where the intertwining kernels are expressed through
dequantizers and quantizers as follows:
\begin{eqnarray}
\label{intertw-kernels} && \fl K_{\mathfrak{F} \rightarrow
\mathfrak{N}} (m,n;m',n') = {\rm Tr} \big[
\hat{U}^{(\mathfrak{N})\dag}(m,n) \hat{D}^{(\mathfrak{F})}(m',n')
\big] = \frac{\sqrt{m'!n'!}}{(n'-m)!} \delta_{m+m',n+n'},\\
&& \fl K_{ \mathfrak{N} \rightarrow \mathfrak{F}} (m',n';m,n) =
{\rm Tr} \big[ \hat{U}^{(\mathfrak{F})\dag}(m',n')
\hat{D}^{(\mathfrak{N})}(m',n') \big] = \frac{(-1)^{m-n'}
\delta_{m+m',n+n'}}{(m-n')!\sqrt{m'!n'!}}.
\end{eqnarray}

It is worth mentioning that the dequantizers and quantizers of a
star-product scheme should not necessarily depend on discrete
variables. For example, in optical tomography, the dequantizer is
a projection on the rotated quadrature, i.e.
$\hat{U}^{(\mathfrak{O})}(X,\theta) = |X,\theta\rangle \langle
X,\theta |$, where $(\hat{q} \cos\theta  + \hat{p} \sin\theta
)|X,\theta\rangle = X |X,\theta\rangle$, $\hat{q}$ and $\hat{p}$
are the position and momentum operator, respectively. The
intertwining formulas that connect the optical tomogram and the
normally (antinormally) ordered moments are derived in the
papers~\cite{richter,wunsche-96,filippov}.

\section{\label{section-evolution} Evolution equations}

As it is outlined in \Sref{introduction}, the star-product
formalism is quite useful for describing evolution of quantum
states. In case of the star-product scheme
(\ref{dequantization})--(\ref{quantization-2}), one deals with the
functions of discrete variables instead of operators. The more
important fact is that the functions $f_{\rho}(m,n;t)$ are
experimentally measurable~\cite{menzel,mariantoni,eichler} at
different time moments $t$, giving an opportunity to observe
dynamics of the system and motivating us to derive the evolution
equations in terms of measurable quantities.

To start with, the time dependent and stationary von Neumann
equations for the density operator $\hat{\rho}$ take the following
form within the star-product formalism:
\begin{eqnarray}
\label{vN-symbols} && \fl \frac{\partial \hat{\rho}}{\partial t} =
- i \big[ \hat{H}, \hat{\rho} \big] ~ \Leftrightarrow ~
\frac{\partial f_{\rho}(m,n;t)}{\partial t} = - i \left( f_H \star
f_{\rho} - f_{\rho} \star
f_H \right)(m,n;t), \\
\label{vN-stationary-symbols} && \fl \frac{1}{2} \left( \hat{H}
\hat{\rho}_E + \hat{\rho}_E \hat{H} \right) = E \hat{\rho}_E ~
\Leftrightarrow ~ \frac{1}{2} \left( f_H \star f_{\rho_E} +
f_{\rho_E} \star f_H \right)(m,n) = E f_{\rho_E} (m,n),
\end{eqnarray}

\noindent where $\hat{H}$ and $E$ are the Hamiltonian and the
permitted energy level, respectively; the Planck constant
$\hbar=1$.

\subsection{Moments' dynamics for harmonic oscillator}

Let us consider a free evolution of the electromagnetic field
governed by a harmonic oscillator Hamiltonian $\hat{H} =
\hat{a}^{\dag}\hat{a} + \textstyle\frac{1}{2}$, where we have also
used dimensionless units for the frequency ($\omega = 1$). In this
case we immediately encounter a problem of finding the symbol
$f_{H}(m,n)$ because it takes infinite values if $m=n$. However,
this difficulty can be avoided if we focus on the star product of
symbols. In fact,
\begin{eqnarray}
\fl \left(f_H \star f_{\rho}\right)(m,n) &=& {\rm Tr} \big[
(\hat{a}^{\dag})^m \hat{a}^n \big( \hat{a}^{\dag}\hat{a} +
\textstyle\frac{1}{2} \big) \hat{\rho} \big] = {\rm Tr} \big[
(\hat{a}^{\dag})^m \big( \hat{a}^n \hat{a}^{\dag} \big) \hat{a}
\hat{\rho} \big] + \textstyle\frac{1}{2} {\rm Tr} \big[
(\hat{a}^{\dag})^m
 \hat{a}^n  \hat{\rho} \big] \nonumber\\
&=& {\rm Tr} \big[ (\hat{a}^{\dag})^m \big( \hat{a}^{\dag}
\hat{a}^n + n \hat{a}^{n-1} \big) \hat{a} \hat{\rho} \big] +
\textstyle\frac{1}{2} {\rm Tr} \big[ (\hat{a}^{\dag})^m
 \hat{a}^n  \hat{\rho} \big] \nonumber\\
&=& f_{\rho}(m+1,n+1)
 + \big( n + \textstyle\frac{1}{2} \big) f_{\rho}(m,n).
\end{eqnarray}

Similarly, we obtain $\left(f_{\rho} \star f_H\right)(m,n) =
f_{\rho}(m+1,n+1)
 + \big( m + \textstyle\frac{1}{2} \big) f_{\rho}(m,n)$.
Substituting these results in Eqs.
(\ref{vN-symbols})--(\ref{vN-stationary-symbols}) yields
\begin{eqnarray}
&& \label{harmonic-evolution} \frac{\partial
f_{\rho}(m,n;t)}{\partial t} =
i (m-n) f_{\rho}(m,n;t),\\
&& \label{harmonic-energy} f_{\rho_E}(m+1,n+1) + \frac{m+n+1}{2}
f_{\rho_E}(m,n) = E f_{\rho_E}(m,n).
\end{eqnarray}

It is not hard to check that symbols $f_{|N\rangle\langle
N|}(m,n;t) = N! e^{i(m-n)t} \delta_{m,n} / (N-m)!$ of the Fock
states $|N\rangle$ such that $m=n\le N$ do satisfy the derived
equations (\ref{harmonic-evolution})--(\ref{harmonic-energy}) if
$E=N+\textstyle\frac{1}{2}$.

\begin{figure}
\begin{center}
\includegraphics[width=10cm]{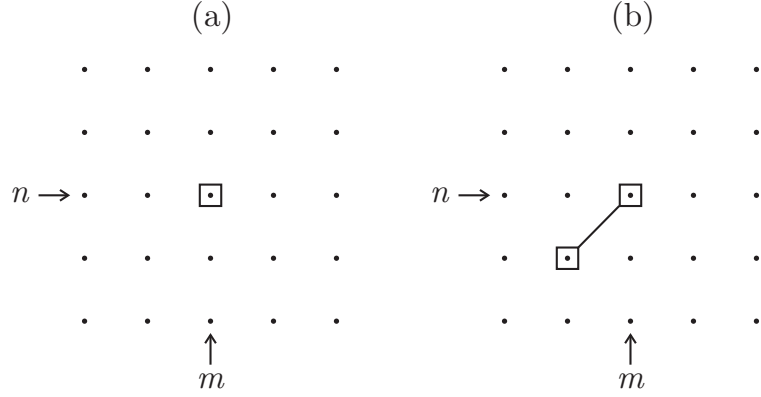}
\caption{\label{figure-lattice-harmonic} Local dynamics of the
moment $\langle (\hat{a}^{\dag})^m \hat{a}^n \rangle$ on the
``phase-space'' lattice: (a) harmonic oscillator and (b) damped
harmonic oscillator. Squares denote nodes for which the
corresponding moments are involved in the evolution of $\langle
(\hat{a}^{\dag})^m \hat{a}^n \rangle$.}
\end{center}
\end{figure}

In case of the harmonic oscillator Hamiltonian, the dynamics
(\ref{harmonic-evolution}) of quantum states on the lattice
$(m,n)$ reduces to $f_{\rho}(m,n;t) = f_{\rho}(m,n;0)
e^{i(m-n)t}$, i.e. the moments simply gain phases in accordance
with their position on the lattice. The corresponding vectors in
Fig.~\ref{figure-phase-space}b save their length and rotate with
frequencies $\omega_{mn} = m-n$, so in time interval $t=2\pi$ all
the vectors come back to the initial position. Thus, evolution of
the moment $\langle (\hat{a}^{\dag})^m \hat{a}^n \rangle$ is
extremely local on the ``phase-space'' lattice and does not depend
on the values of moments in surrounding nodes
(Fig.~\ref{figure-lattice-harmonic}a).

\subsection{Moments' dynamics for damped harmonic oscillator}

Let us consider a damped evolution equation for the density
operator described by the usual master equation (see, e.g.,
\cite{scully,dodonov-mizrahi-souzasilva})
\begin{equation}
\label{damped-rho} \fl \frac{\partial \hat{\rho}}{\partial t} = -i
[\hat{a}^{\dag}\hat{a}, \hat{\rho}] + \gamma (1+\nu) (2 \hat{a}
\hat{\rho} \hat{a}^{\dag} - \hat{a}^{\dag}\hat{a} \hat{\rho} -
\hat{\rho} \hat{a}^{\dag}\hat{a}) + \gamma \nu (2 \hat{a}^{\dag}
\hat{\rho} \hat{a} - \hat{a}\hat{a}^{\dag} \hat{\rho} - \hat{\rho}
\hat{a}\hat{a}^{\dag}),
\end{equation}

\noindent where $\gamma$ is the damping coefficient and $\nu$ is
the equilibrium mean number of photons in a given mode.

Arguing as above, we find the star product of symbols in question
\begin{eqnarray}
\fl \left(f_a \star f_{\rho} \star f_{a^{\dag}} \right)(m,n) &=&
f_{\rho}(m+1,n+1),\\
\fl \left(f_{a^{\dag}} \star f_{\rho} \star f_a \right)(m,n) &=&
f_{\rho}(m+1,n+1) + (m+n+1) f_{\rho}(m,n) \nonumber\\
&& + m n f_{\rho}(m-1,n-1).
\end{eqnarray}

Now one can rewrite (\ref{damped-rho}) in terms of the measurable
moments $\langle (\hat{a}^{\dag})^m \hat{a}^n \rangle \equiv
f_{\rho}(m,n)$ as follows:
\begin{equation}
\label{damped-symbols} \fl \frac{\partial
f_{\rho}(m,n;t)}{\partial t} = \left[ i(m-n) - \gamma (m+n)
\right] f_{\rho}(m,n;t) + 2 \gamma \nu m n f_{\rho}(m-1,n-1;t).
\end{equation}

In case of the damped harmonic oscillator, the dynamics of moments
is again quite local on the lattice $(m,n)$ and involves only two
nodes (Fig.~\ref{figure-lattice-harmonic}b). It is worth
mentioning, that Eq. (\ref{damped-symbols}) could also be derived
by using a connection between the normally ordered moments and the
Wigner function, and then substituting these relations in the
Fokker-Planck equation for the Wigner function \cite{filippov}.
However, as we can see, the star-product approach is
straightforward to derive evolution equations for measurable
quantities like the normally ordered moments of the creation and
annihilation photon operators.

\subsection{Moments' dynamics for a particle}

\begin{figure}
\begin{center}
\includegraphics[width=16cm]{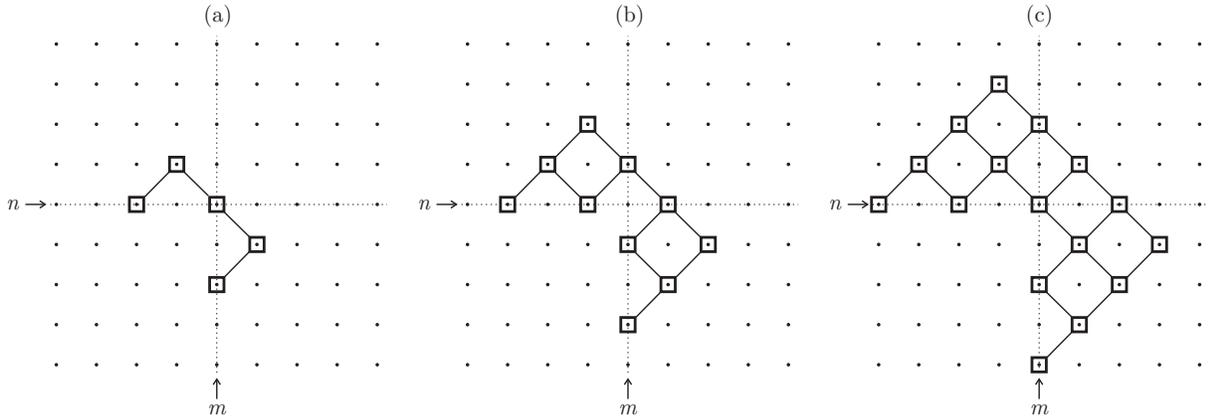}
\caption{\label{figure-lattice-particle} Nodes on the
``phase-space'' lattice for which the corresponding moments are
involved in the evolution of a particle: (a) kinetic term
$\hat{p}^2/2$, $V(\hat{q}) \propto \hat{q}^2$; (b) $V(\hat{q})
\propto \hat{q}^3$; (c) $V(\hat{q}) \propto \hat{q}^4$. The lower
sum $m+n$ for the marked nodes the greater weight of corresponding
moments in the time evolution.}
\end{center}
\end{figure}

Although the background of our consideration is measuring
radiation fields at microwaves, the developed star-product
formalism can be also applied to the one-dimensional motion of
particles governed by the Hamiltonian $\hat{H}=\hat{p}^2/2 +
V(\hat{q}) = - (\hat{a}-\hat{a}^{\dag})^2/4 +
V((\hat{a}+\hat{a}^{\dag})/\sqrt{2})$. The main point is that the
time evolution of moments remains local on the lattice if the
potential energy $V(q)$ can be approximated by lower-order terms
of the Taylor expansion (Fig.~\ref{figure-lattice-particle}).

If $V(q)=q^l$, then the nodes $(m',n')$ involved in the dynamics
of the moment $\langle (\hat{a}^{\dag})^m \hat{a}^n \rangle$ form
a truncated square with center at the point $(m-1,n-1)$. The nodes
$(m',n')$ satisfy the relations
\begin{eqnarray}
&& |m'-m+1| + |n'-n+1| = \left\{ \begin{array}{c}
  0,2,4,\ldots,l \quad {\rm for ~ even }~l, \\
  1,3,5,\ldots,l \quad {\rm for ~ odd }~l, \\
\end{array}\right. \nonumber\\
&& \max(m'-m,n'-n) \ge 0. \nonumber\\
\end{eqnarray}

It is instructive to compare the $W$-function evolution and the
evolution of moments. The derivatives $d^{l} V/ d q^{l}$ and
powers $q^l$ form the Taylor expansion of $V(q)$. These
derivatives show which partial derivatives $\partial^{l}W/\partial
p^{l}$ contribute to the evolution on the conventional phase space
$(q,p)$, whereas the powers $q^l$ show which moments $\langle
(\hat{a}^{\dag})^m \hat{a}^n \rangle$ contribute to the evolution
on the ``phase-space'' lattice $(m,n)$. The important practical
difference between a quasidistribution on the conventional phase
space $(q,p)$ and our proposal is that the time development of a
conventional quasidistribution is given in terms of the
infinite-order partial differential equation which is not so easy
to solve. A numerical solution of such an equation would require a
construction of a two-dimensional rectangular grid in the
$(q,p)$-plane, with the size and density of the grid being taken
according to a desired accuracy (see, e.g.,~\cite{lee}). The
partial derivatives are then replaced by finite differences. The
higher the order of the derivative, the more nodes of the grid are
involved. Dealing with the ordered moments, one does not need to
introduce any artificial grid because of the lattice itself. The
time development is then given by the exact difference equations
in contrast to the approximate finite-difference equations for
quasidistributions on the $(q,p)$-plane.

\section{\label{section-conclusions}Conclusions}
To conclude, we present the main results of the paper.

An analysis of the star-product scheme based on the normally
ordered creation and annihilation photon operators has been
motivated by the recent advances in measuring ordered moments for
microwave quantum states \cite{menzel,mariantoni,eichler}. In
addition, the phase space of the constructed quantization scheme
is a two-dimensional lattice $(m,n)$ whose nodes are given by two
nonnegative integers. Such a structure of the phase space is
advantageous for describing the time development of some quantum
systems because the exact evolution equations take the form of
difference equations in contrast to the partial differential
equations for conventional quasidistributions on the $(q,p)$-plane
usually approximated by finite-difference equations on the
$(q_i,p_j)$-grid of rather artificial size and density.

Moreover, it is quite reasonable to define a quantum state
evolution in terms of the measurable quantities, so we have filled
a gap of such equations in terms of the normally ordered moments
$\langle (\hat{a}^{\dag})^m \hat{a}^n \rangle$. The dynamics of
moments is shown to be local on the lattice for (damped) radiation
fields and particles moving in smooth potentials.

Another substantial result is that the star-product kernel is
found in the explicit form (\ref{kernel-explicit}). As any
star-product kernel is to satisfy specific non-linear equalities,
we have applied one of those equalities to the obtained kernel and
thus derived new sum relations involving factorials
(\ref{sum-rel-1})--(\ref{sum-rel-4}).

\ack The authors thank the Russian Foundation for Basic Research
for partial support under Projects Nos. 09-02-00142, 10-02-00312,
and 11-02-00456. S.N.F. is grateful to the Dynasty Foundation and
the Russian Science Support Foundation for support under Project
``Best postgraduates of the Russian Academy of Sciences 2010".
S.N.F. thanks the Ministry of Education and Science of the Russian
Federation for support under Project Nos. 2.1.1/5909, $\Pi$558,
and 14.740.11.1257.

\end{document}